\documentclass[aps,prx,twocolumn,groupedaddress]{revtex4-1}

\usepackage{amsmath}
\usepackage{amssymb}
\usepackage{graphicx}
\usepackage{anysize}
\usepackage{color}
\newcommand{\R}{\ensuremath{\mathbb{R}}}

\graphicspath{{./}{Figs/}}

\begin{document}

\title{
A Kullback-Leibler divergence measure of intermittency: \\
application to turbulence
}
\author{Carlos Granero Belinch\'on}
\author{St\'ephane G. Roux}
\author{Nicolas B. Garnier}
\affiliation{Univ Lyon, Ens de Lyon, Univ Claude Bernard, CNRS UMR 5672, Laboratoire de Physique, F-69342 Lyon, France
}

\date{\today}

\begin{abstract}

For generic systems exhibiting power law behaviors, and hence multiscale dependencies,
we propose a new, and yet simple, tool to analyze multifractality and intermittency, after noticing that these concepts are directly related to the deformation of a probability density function from Gaussian at large scales to non-Gaussian at smaller scales.
Our framework is based on information theory, and uses Shannon entropy and Kullback-Leibler divergence. 

We propose an extensive application to three-dimensional fully developed turbulence, seen here as a paradigmatic complex system where intermittency was historically defined. 
Moreover, the concepts of scale invariance and multifractality were extensively studied in this field and, most importantly, benchmarked.
We compute our measure on experimental Eulerian velocity measurements, as well as on synthetic processes and a phenomenological model of fluid turbulence.

Our approach is very general and does not require any underlying model of the system, although it can 
probe the relevance of such a model. 

\end{abstract}

\maketitle 


\section{Introduction}

Complex systems are omnipresent in day life, and as a consequence in many scientific fields. 
These are as various as:
internet traffic~\cite{Feldmann1998,Fontugne2017}, human genome exploration~\cite{Boulos2017}, geography~\cite{Ahn2010}, financial markets~\cite{Bouchaud2003,Borland2005}, {\em etc}.
In recent years, increasing computational power and storage has fuelled interest in accumulating and analyzing large amounts of data, and hence in developing new tools to characterize systems where traditional methods are not relevant. 
Real world systems are most commonly nonlinear and their complexity is difficult to model. Hence, any new tool must be able to probe non-linear correlations and should preferably be non-parametric.

A very common characteristic of such systems is the occurence of quantities that
exhibit a power spectral density (PSD) with a power law, indicating that multiple scales are present, in a continuous range.
In addition, probability density functions (pdfs) are most commonly non-Gaussian, suggesting that non-linear interactions are at work.
When one defines and then measures a global or local quantity which has both a power law power spectrum and a non-Gaussian pdf, this quantity is usually shown to have a fractal nature, with a deformation of its pdf when the scale varies.
Again, very few tools exist to probe such systems correctly.
We propose here a novel measure of the evolution between Gaussian and non-Gaussian pdf.
Another wide class of problems consider a global quantity defined as the sum or integral, e.g., over space or over time, of a local quantity that has non-Gaussian statistics. 
Estimating or predicting the statistics of the global quantity is usually not easy, because of, e.g., long-range interactions that leads to long-range correlations.
Nevertheless, for large integration scales ---~larger than any possible correlation scale~---, large deviation theory should reduce to central limit theorem and Gaussian statistics are expected. 
We also aim at quantifying these transformations of the pdfs.

\medskip
One paradigm of multiscale complex systems is fluid turbulence, for which a lot of theoretical and phenomenological developments have been proposed to describe very complex behaviours such as energy cascade and intermittency.

Kolmogorov 1941 theory (K41)~\cite{Kolmogorov1941a,Kolmogorov1941b,Kolmogorov1941c} provided a powerful framework to describe fully developed turbulence, and in particular to characterize its statistical properties.
Considering for example Eulerian longitudinal velocity increments $\delta_l v$ at scale $l$, 
K41 theory states that the $p$-order structure function $S_{p}(l) \equiv \left\langle \left( \delta_l v\right)^p \right\rangle$ behaves, for any positive integer $p$, as a power law of the scale $l$:
\begin{equation}
S_{p}(l) 
\propto l^{\zeta(p)}
\label{eq:structure_function}
\end{equation}
in the inertial range: $\eta \ll l \ll L$, where $L$ and $\eta$ are the integral and Kolmogorov scales. 
The scaling exponent $\zeta(p)$ depends on the order $p$, and 
K41 theory, assuming homogeneity and isotropy at the small scales of the flow, 
predicts a linear behavior of the scaling exponents as $\zeta(p)=p/3$. 
Using the relation between the kinetic energy and the second order structure function $S_{2}(l)$, 
this implies the famous 5/3 law for the distribution of kinetic energy in the inertial range.
The existence of the energy cascade from larger scales to smaller ones has been shown by Kolmogorov to be related to the 4/5 law, which imposes the value of the third order structure function
$S_{3}(l)$.

Although very satisfying at first, this predicted linear behavior of the scaling exponents was later rejected by dedicated experiments~\cite{F.Anselmet1984}.
To describe the observed deviations, Kolmogorov and Oboukhov relaxed some of the K41 hypothesis 
and in particular the homogeneity at small scale~\cite{Kolmogorov1962,Oboukhov1962}. 
This led to the KO62 theory and the definition of intermittency in Turbulence: the scaling exponents $\zeta(p)$ do not depend linearly in $p$ (see Figure~\ref{fig:synth:1}).

Intermittency was later related to the multifractal description of turbulence as stated by Frisch and Parisi~\cite{Frisch1985}.  
It is now recognized that the pdf of the Eulerian velocity increments continuously deforms from Gaussian at large scales $l \gtrsim L$ to strongly non-Gaussian at smaller scales, and we use this consequence as an equivalent, but more practical, definition of intermittency.

\medskip

The deformation of the pdf can be quantified by the evolution of its flatness~\cite{Frisch1995}, measured as the normalized kurtosis of the distribution: $\langle (\delta_l v)^4\rangle / \langle (\delta_l v)^2\rangle^2 = S_4(l)/S_2(l)^2$.
%
%
At larger scales, about or above the integral scale $L$, the pdf of the velocity increments is Gaussian and therefore has a flatness equal to $3$. For smaller and smaller scales, the pdf is less and less Gaussian as the pdf of the normalized increments becomes wider and wider; therefore the flatness increases. Fine evolutions of the pdf, and hence intermittency, have been studied with the flatness, such as a rapid increase of intermittency when the scale is reduced down or below the Kolmogorov dissipative scale~\cite{Chevillard2005}. 
$S_2$ evolves according to the 5/3 law, so, the kurtosis only involves one higher-order structure function, namely $S_4$, and as such it does not describe the deviation of all the scaling exponents $\zeta(p)$ from their linear behavior in $p$. This is why we propose in this article a new measure of intermittency that involves all structure functions.


Shannon founded Information theory in 1948 and introduced entropy as a measure of the total information of a process~\cite{Shannon1948}. 
Since then, Information Theory has been widely used in very different fields: biomedical science \cite{Pincus1992,Porta2001,Porta2013}, Physics of fluids \cite{Brown1982,Ikeda1989,Cerbus2013}, thermodynamics~\cite{J.M.Horowitz2014} and others.  
Shannon entropy is a functional of the pdf of the process, and as such it depends on all the moments of the distribution, see section \ref{Sec:def}. 


We provide a new measure of intermittency, interpreted as the deformation of a pdf which is Gaussian at large scales. 
To do so, we consider the Kullback-Leibler (KL) divergence~\cite{Kullback1968} between the pdf and the Gaussian pdf defined as having the same standard deviation. 
We compute this quantity for Eulerian velocity increments, in order to measure intermittency in turbulence.
We do so in a wide range of scales so that we can observe how this measure of intermittency behaves in all domains of fluid turbulence. 
By comparing the pdf --- defined by all its moments --- and the Gaussian approximation of this pdf --- defined by the second order moment only --- we measure not only the growth of the $p$th order moment with respect to the variance, but also the evolution of all the moments with respect to the variance, {\em i.e.}, we exhaustively characterize the deformation of the pdf.
Measuring the intermittency with a KL divergence provides a generalization of measures such as flatness ($p=4$), hyperflatness ($p=6$), {\em etc}. 

\medskip

Although we propose to study turbulence as an application of our framework, our definitions are very general, and only require a signal to probe intermittency. Our approach does not require any {\em a priori} knowledge of the signal, neither any underlying model of the system that produced the signal. As such, it can prove a very powerful tool to analyze complex systems exhibiting power law behaviors or multiscale dependencies.

This paper is organized as follow. In section~\ref{Sec:def}, we define our information theoretical measure of intermittency that involves Shanon entropy and a well chosen Kullback-Leibler divergence. In section~\ref{sec:Turbulence}, we compute this quantity for experimental measurements of the Eulerian velocity field in several setups and several Reynolds numbers. We then turn in section~\ref{sec:modelling} to some phenomenological modelings in order to better understand and describe our observations.

\section{Definitions} 
\label{Sec:def}

\subsection{Entropy and KL Divergence from Gaussianity}

Shannon entropy,  $H(X)$, of a process $X$ of pdf $p(x)$ is the total information that defines the process \cite{Shannon1948}. It depends on all the moments of the pdf $p(x)$ except the first order one. 

\begin{equation} \label{eq:H}
H(X)=- \int_{\R} p(x)\log p(x) {\rm d}x.
\end{equation}

We know that a Gaussian process, $X_{G}$, is defined only in terms of two-point correlations. 
Therefore, its Shannon entropy only depends on its variance $\sigma_{X_{\rm G}}$, and we have the analytical expression of the entropy $H(X_{G})$ of a Gaussian process $X_G$:
\begin{equation} \label{eq:HG}
 H(X_{\rm G}) = \frac{1}{2}\log(2 \pi e \sigma_{X_{\rm G}}^{2}) \,.
\end{equation}

For a generic process $X$ which is {\em a priori} non-Gaussian and has the variance $\sigma_X^2$, we define the ``entropy under Gaussian hypothesis'' $H_{\rm G}(X)$ as the entropy that one would get assuming the process is Gaussian and using eq.(\ref{eq:HG}):
\begin{equation} \label{eq:HG2}
 H_{\rm G}(X) = \frac{1}{2}\log(2 \pi e \sigma_{\text{\tiny{$X$}}}^{2}) \,,
\end{equation}
where $\sigma_{\text{\tiny{X}}}$ is the standard deviation (std) of the generic process $X$.
So, the ''entropy under Gaussian hypothesis'' of $X$ is a measure of the entropy of a Gaussian pdf with  same std as the real pdf of $X$. 
If $X$ is Gaussian, obviously $H_{\rm G}(X_{\rm G})=H(X_{\rm G})$. 

\medskip

For any process $X$ with probability density function $p(x)$, we can measure the difference between the "real" pdf $p(x)$ of $X$ and the Gaussian approximation $p_{\rm G}(x)$ using the Kullback-Leibler divergence~\cite{Kullback1968}:
\begin{equation}\label{eq:KL}
{\cal K}_{p||p_{\rm G}}(X)=\int_{\R} p(x)\log \left( \frac{p(x)}{p_{\rm G}(x)} \right) {\rm d}x  \,.
\end{equation}
Using the definitions of $H(X)$ and $H_G(X)$, we have:
\begin{equation}\label{eq:KL2}
{\cal K}_{p||p_{\rm G}}(X)= H_{\rm G}(X)-H(X)\geq0 \,.
\end{equation}

${\cal K}_{p||p_{\rm G}}(X)$ is a measure of the distance from Gaussianity of the process $X$, {\em i.e.}, the distance between the pdf $p(x)$ of $X$, and a Gaussian pdf $p_{\rm G}(x)$ which has the same std. 
The maximum entropy principle~\cite{Jaynes1957,Jaynes1957a} states that for a given standard deviation, the Gaussian pdf maximizes the entropy, 
see also~\cite{cover2006}. So this distance is also a comparison between the total information needed to define the 
process and the total information defining the most ambiguous process with same std. 
The maximization of the entropy for the Gaussian case ensures that the difference $H_{\rm G}(X)-H(X)$ is always positive, as expected for a KL divergence and vanishes only when $X$ has a Gaussian distribution.

\subsection{Distance from Gaussianity across scales}

We analyse the process $X$ at scale $\tau$ by studying its increments of size $\tau$:
\begin{equation}\label{eq:incr}
\delta_{\tau}X(t)=X(t+\tau)-X(t) 
\end{equation}

We note $D_{\tau}(X)$ the KL divergence ${\cal K}_{p||p_{\rm G}}(\delta_{\tau}X)$ which measures the distance from Gaussianity of the increments at scale $\tau$ of a process $X$:

\begin{equation} \label{eq:dist}
D_{\tau}(X)={\cal K}_{p||p_{\rm G}}(\delta_{\tau}X)= H_{\rm G}(\delta_{\tau}X)-H(\delta_{\tau}X)
\end{equation}

This quantity measures the deformation of the pdf of the increment as a function of the size $\tau$ of the increment: it quantifies the evolution of the shape of the pdf, which depends on all the moments of the process, except its mean.

Indeed, at each scale $\tau$, the increment $\delta_{\tau}X$ has a different standard deviation. The larger the scale $\tau$ the higher the standard deviation. 
So, changing $\tau$ changes quantitatively the entropies $H(\delta_{\tau}X)$ and $H_G(\delta_{\tau}X)$, which both depend strongly on the standard deviation. 
Substracting the two entropies eliminates most of this quantitative variation because the std is by construction the same in both expressions $H_G(\delta_{\tau}X)$ and $H(\delta_{\tau}X)$).
$D_{\tau}(X)$ thus only measures subtle and delicate evolutions of the shape of the pdf than the trivial rescaling induces by the std.

In the specific case of Turbulence, the pdfs of the increments of size equal or larger than the integral scale $L$ are almost Gaussian. As a consequence, we expect that the distance from Gaussianity $D_{\tau}(X)$ tends to zero when $\tau$ approaches the integral scale.
Conversely, it is expected to increases in the inertial range down to the dissipative scale where it should increase (even) faster~\cite{Chevillard2005}. 
Our distance from Gaussianity should therefore be able to probe 
intermittency of Turbulence by measuring the deformation of the pdf of velocity increments.   

\subsection{Methodology}

To compute the Shannon entropy $H$, we use a nearest neighbors estimator described by Kozachenko and Leonenko~\cite{L.Kozachenko1987,Leonenko2008}. The only parameter used in this algorithm is the number of neighbors $k$ involved in the nearest neighbors search.
We chose the usual value $k=5$ which is large enough to estimate the Shannon entropy correctly within a reasonable computational time.

Following Theiler~\cite{Theiler1986}, we subsample the data in order to remove spurious correlation effects: when computing the entropy of $\delta_{\tau}X(t)$, we only retain data points separated in time by a delay time $\tau_{max}$, defined as the size of the largest increment that we compute. 
This prescription has two benefits.
Firstly, the correlation between two successive points of the subsampled dataset are uncorrelated, because increments of size $\tau$ are typically correlated over a time $\tau \le \tau_{\rm max}$.
Secondly,  the number $N$ of points used in the computation of the entropy of $\delta_\tau X$ is independent of $\tau$, so the bias due to finite size effects is constant when $\tau$ is varied.

To compute the entropy under Gaussian hypothesis $H_{\rm G}$, we estimate the standard deviation of the process and then use eq.(\ref{eq:HG}). 

In the remainder of this article, all quantities are computed using $N=512$ points, $\tau_{\rm max}=4096$, so signals with a total of $N\tau_{\rm max}=2^{21}$ points. We also average our results over  independent realisations, in order to compute the standard deviation of the quantities and provide error bars to the estimations. We use $12$ realisations for experimental signals and $8$ realisations for synthetic processes.

\medskip

In following sections we analyse the evolution of the pdf along the scales for a longitudinal turbulent velocity signal. We compare the obtained results with some synthetic and theoretical models of turbulence.

\section{Turbulence}
\label{sec:Turbulence}

\subsection{Experimental signals}

We analyse two different sets of experimental turbulent data, in order to show the ability of our measures to grasp inherent properties of turbulence.

The first system consists of a temporal measurement of the longitudinal velocity ($V$) at one location in a grid Turbulence setup in the wind tunnel of ONERA at Modane~\cite{Castaing1998}. 
The Taylor-scale based Reynolds number$\mathbb{R}_{\lambda}$ is about 2700, with a turbulence rate about $8\%$. The inertial region length is approximately three decades. The sampling frequency is $f_s=25$ kHz and the mean velocity of the wind in the tunnel is $\left\langle v \right\rangle=20.5$ m/s. 
The probability density function of the data is almost Gaussian although there is some visible asymmetry: the skewness is about $0.175\pm0.001$.

The second system is a set of temporal velocity measures at different Reynolds numbers in a jet turbulence experiment with Helium~\cite{Chanal2000}. The Taylor-scale based Reynolds number $\mathbb{R}_{\lambda}$ is respectively 89, 208, 463,703,929, with a turbulence rate about 23\%.

Using Taylor hypothesis~\cite{Frisch1995} and the mean 
velocity $\langle v \rangle$ of the flow, we can interpret these time series as the spatial evolution of the longitudinal velocity. 
The time scale $\tau$ and the spatial scale $l$ are related by $l =  \langle v \rangle \tau$.
We note the integral time scale $T$ and the integral spatial scale $L$, and we have $L =  \langle v \rangle T$.
We present all our results as functions of the ratio $\tau/T = l/L$ between the scale of the increment and the integral scale.


\subsection{Results}

\begin{figure}
\centering
\includegraphics[width=\linewidth]{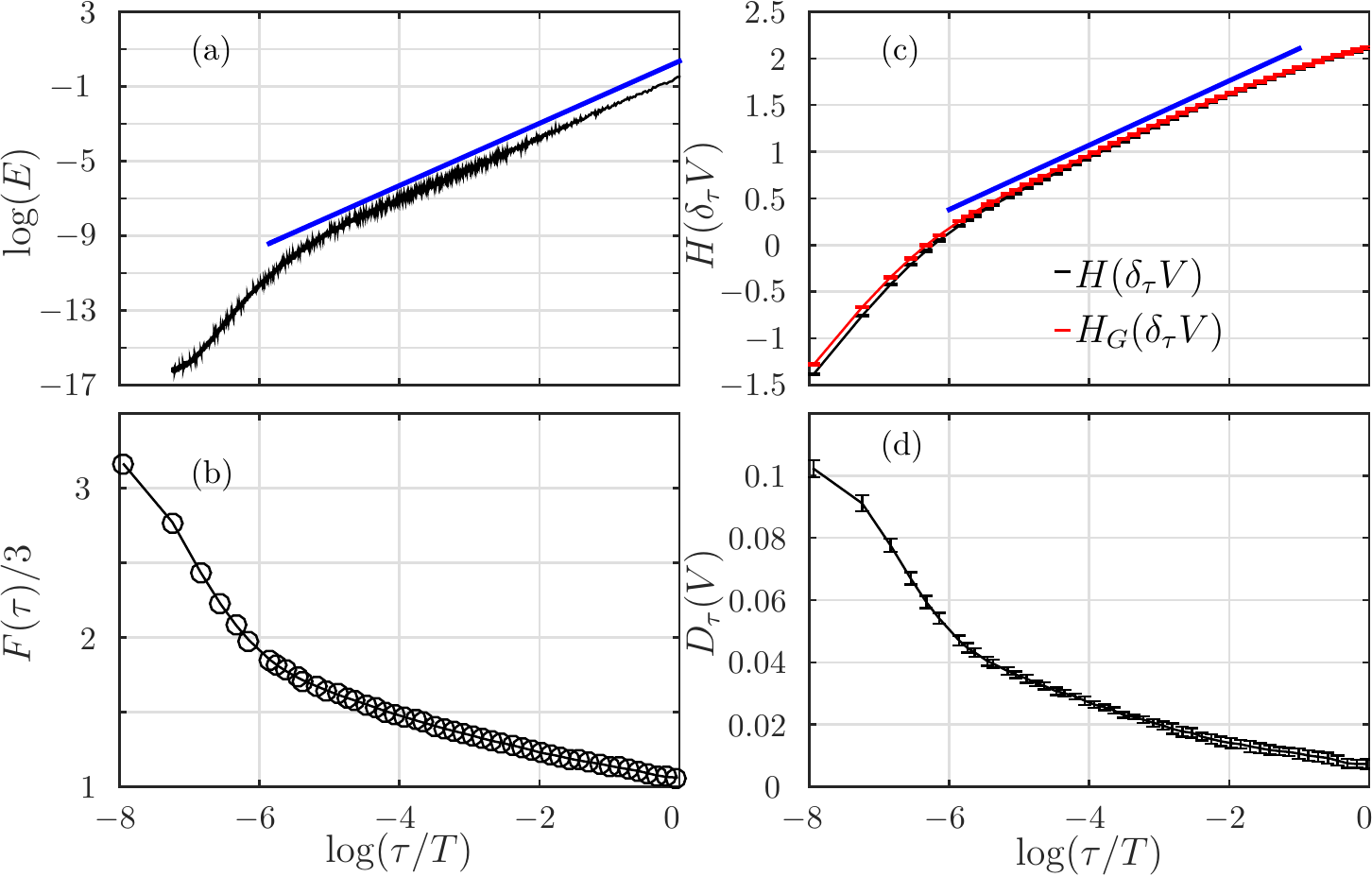}
\caption{(a) Power spectrum (b) Flatness (c) Entropies (d) KL Distance from Gaussianity, for the Modane experimental data, as functions of $\log(\tau/T)=\log(l/L)$, the logarithm of scale normalized by the integral scale.
In (a) and (c), the straight lines indicate the theoretical scaling in the inertial region predicted by Kolmogorov $5/3$ law. 
}
\label{fig:turb:2}
\end{figure}

In Figure~\ref{fig:turb:2}, we present the analysis of the Modane experimental velocity data. 
In the left column we report the classical viewpoint and compare it to the information theory viewpoint in the right column. 
%

We first plot the power spectrum $S(\tau)$ of the velocity signal $V$ in Fig.~\ref{fig:turb:2},a : it shows the distribution of energy across scales following the well known $5/3$ Kolmogorov law. 
In order to measure the deformation of the shape of the pdf of the velocity increment when the scale $ \tau$ is varied, we follow Frisch~\cite{Frisch1995}, and compute the flatness of the velocity increments normalised by the flatness of a Gaussian pdf ($\frac{F(\tau)}{3}$). Results are reported in~\ref{fig:turb:2}(b). 
For $\tau \gtrsim T$, {\em i.e.}, $l \gtrsim L$, the flatness has the value expected for a Gaussian pdf.
Reducing $\tau$, the flatness increases.
When $\tau$ is smaller than the dissipative scale~\cite{Chevillard2005}, the increase of the flatness is sharper.
Three different regions can be distinguished in both figures: integral, inertial and dissipative.

The right column of Fig.~\ref{fig:turb:2} is devoted to the Information Theory viewpoint on the same characteristics of Turbulence. 
We first plot the entropy of the increments in Fig.~\ref{fig:turb:2}(c) and compare it with the PSD in Fig. a. We then plot $D_\tau(V)$ in Fig.~\ref{fig:turb:2}(d) and compare its behavior in $\tau$ with the flatness.

In figure~\ref{fig:turb:2}(c) we see that the entropy of the increments ($H(\delta_{\tau}V)$) increases with $\tau$. The larger the scale, the higher the total Shannon information needed to completely characterize the increment. 
We can distinguish three different ranges with different dependence of the entropy on the scale.
For the large scales, larger than the integral scale, the entropy reaches its highest value and is then constant. So the most disorganized or complex scales ---~the ones requiring more information to be completely characterized~--- are the scales in the integral domain. 
Within this region, the characterization of the scale does not require more entropy when the size of the increment increases . 
A linear behaviour of the entropy in $\log(\tau/T)$ is found in the inertial region, $\tau \in [10,400]$.
The complexity of the scales, as measured by $H(\delta_{\tau}V)$, decreases linearly in $\log(\tau/T)$ between the integral and the Kolmogorov scales.
For the smallest scales, below the Kolmogorov scale --- which we can measure at $\log(\tau)\approx2$ --- we observe a steeper decrease of the disorganization of the scales with the decrease of 
the scale. 
So, using the entropy of the increments, we are able to recover the three different regions.
Moreover,  we can state that $H(\delta_{\tau}V)$ increases from the smallest scale to the 
integral scale, and then remains constant. In addition, the evolution of $H(\delta_{\tau}V)$ in the inertial region is linear in $\log(\tau/T)$. 

Both entropies $H(\delta_{\tau}V)$ and $H_{G}(\delta_{\tau}V)$ in Fig.~\ref{fig:turb:2},c are indistinguishable in the integral domain. The distance between them starts to increase when we enter in the inertial region. 
In figure~\ref{fig:turb:2},d, we plot the difference between these two entropies, which according to equation(\ref{eq:dist}) is the distance from Gaussianty $D_\tau(V)$.
Starting from 0 at scales larger than the integral one, it increases when the scale decreases. 
The vanishing of $D_\tau(V)$ for largest scales implies that the pdf of the velocity increments is almost Gaussian, which is the expected behavior in Turbulence at scales equal or larger than the integral scale.
Below this integral scale the pdf starts to deform, and becomes less and less Gaussian when the scale decreases. The evolution of $D_\tau(V)$ is almost linear between the integral and the Kolmogorov scales. 
Finally in the dissipative range, we observe 
an abrupt deformation of the pdf, in perfect agreement with the rapid increase of the flatness
in figure~\ref{fig:turb:2},b.~\cite{Chevillard2005}
The distance from Gaussianity $D_\tau(V)$ across scales $\tau$ is a measure of the deformation of the pdf of the turbulent velocity increments, and, as such, a measure of the intermittency.

In the four sub-plots of figure~\ref{fig:turb:2},  the three different domain of turbulence are distinguishable: integral, inertial and dissipative.
Figure~\ref{fig:turb:2}(c) allows us to interpret these three  domains in terms of organization and complexity of velocity increments. 
Figure~\ref{fig:turb:2},d shows that the KL divergence allows us to quantify the evolution of intermittency
amongst scales $\tau$. 
We do not only recover the three different ranges with our new measure based on Information theory, but also the qualitative behavior of intermittency in each domain is in perfect agreement with previous studies.
Moreover, our measure of intermittency doesn't depend on a specific ratio between selected moments of the pdfs like the kurtosis.
$D_\tau(V)$ takes into account all the moments defining the pdfs: this makes our KL distance from Gaussianity across scales a good candidate for a quantitative measure of intermittency.

\begin{figure}[htb]
\centering
\includegraphics[width=\linewidth]{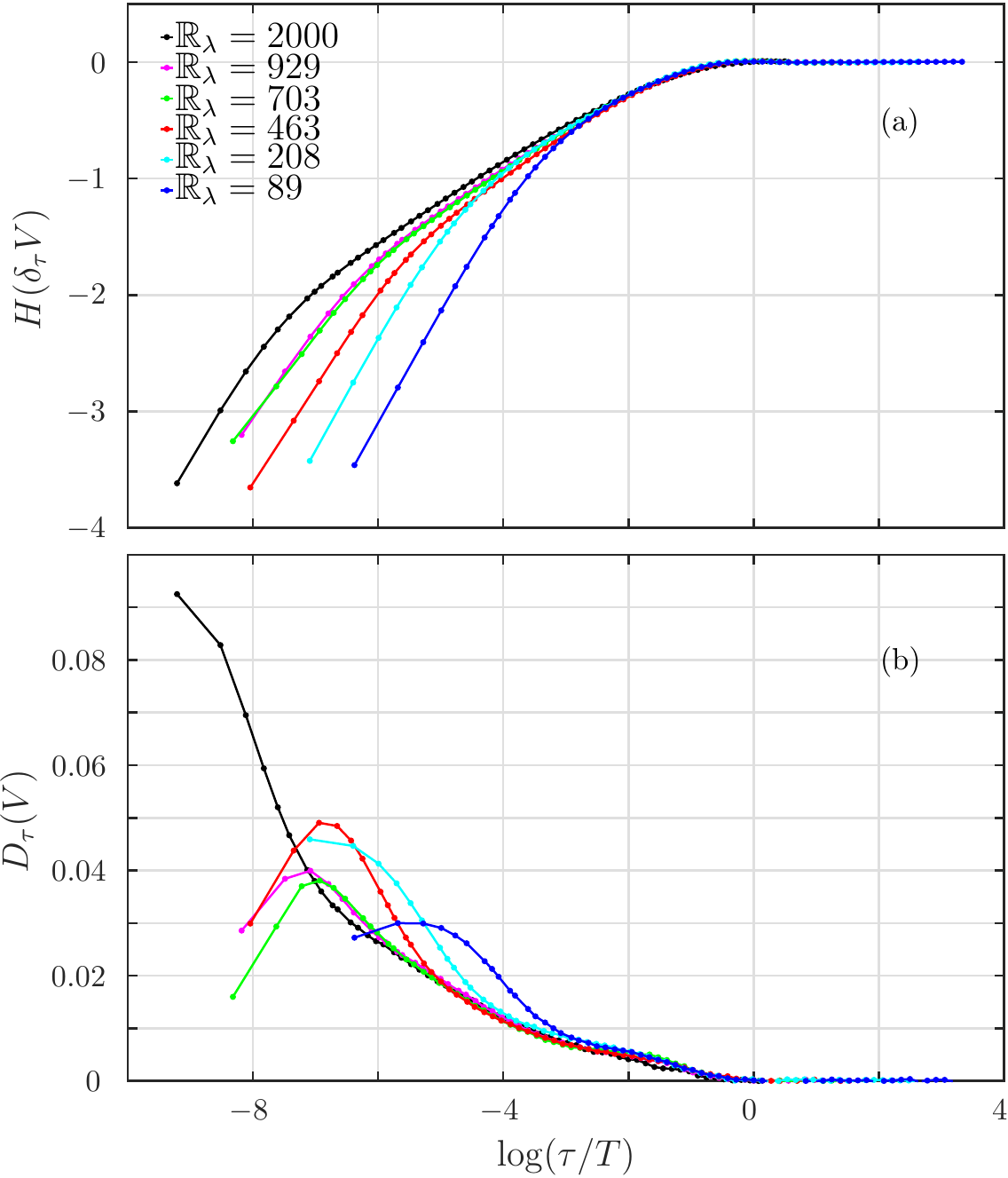}
\caption{a) Entropy $H(\delta_\tau V)$ of the Eulerian velocity increments as a function of $\log(\tau/T)=\log(l/L)$. 
b) KL divergence $D_\tau(V) = H_{\rm G}(\delta_\tau V)-H(\delta_\tau V)$. 
Different experimental signals with various Reynolds numbers have been used.}
\label{fig:turb:RE}
\end{figure}

We have compiled Information Theory results for all experimental signals in Figure~\ref{fig:turb:RE}, in order to study the influence of the Reynolds number.
The entropy as a function of the scale is reported in Figure~\ref{fig:turb:RE}(a); we observe how the size of the inertial range varies with the Reynolds number, with the Kolmogorov scale increasing when the Reynolds number decreases.
This classical behaviour of the Kolmogorov scale is also recovered with the KL divergence, represented in Fig.~\ref{fig:turb:RE}(b). The steeper slope ---~that indicates the dissipative domain~--- appears at higher scales when the Reynolds number is lower; we recover the dependence of the Kolmogorov scales with the Reynolds number. 

The behaviours of both the entropy and the distance from Gaussianity are qualitatively the same for different experimental setups and for any Reynolds number.
The dependence of the entropy $H(\delta_\tau V)$ of the increments is, at first order, in agreement with the K41 theory: we recover the scaling law in the inertial domain~\cite{Granero16}.
The KL divergence $D_\tau(V)$ then enlightens the deformation of the pdf across scales, which is
qualitatively compatible with the K062 theory and hence the intermittency in Turbulence.

\section{Modelling}
\label{sec:modelling}

In order to get some insight on the quantitative results obtained with our Kullback-Leibler divergence $D_\tau$, we now turn to some theoretical descriptions of the inertial domain of fully developed turbulence.

First, we study different processes. Amongst the simplest, popular and most important is fractional Brownian motion (fBm)~\cite{Kolmogorov1941a,M68} which, as a monofractal process, doesn't display intermittency. 
We also explore multifractal processes, which exhibit intermittency:
Multifractal Random Walk (MRW)~\cite{Bacry2001,Leonarduzzi2014}, 
Random Wavelet Cascade (RWC) with Log-normal~\cite{Leonarduzzi2014,Arneodo1998} or Log-Poisson distribution of multipliers~\cite{Leonarduzzi2014,Arneodo1998,She1994}. 
We then examine the propagator formalism~\cite{Castaing1990}, a phenomenological model that provides an analytical expression of the pdf of the velocity increments~\cite{Chevillard2012}.

\subsection{Description of turbulence}

\begin{figure}[htb]
\centering
\includegraphics[width=\linewidth]{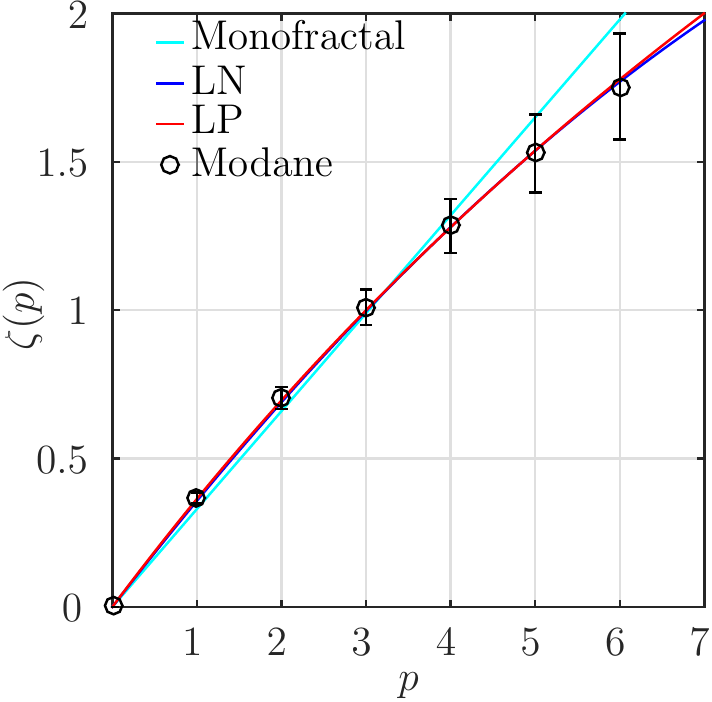}
\caption{Scaling exponents $\zeta(p)$ versus order $p$ for three different models of turbulence in the inertial domain, together with an experimental Eulerian velocity measure (Modane, black).
Models are: monofractal fractional Brownian motion (cyan), multifractal log-normal (blue) and multifractal log-Poisson (red).}
\label{fig:synth:1}
\end{figure}

We briefly introduce the models that we have used.
All are characterized by the set of their scaling exponents, $\zeta(p)$, as they appear in eq.(\ref{eq:structure_function}). 
One of the very few exact results of Kolmogorov's framework is the 4/5-law which
imposes $\zeta(3)=1$; this should be respected by
any model or process representing turbulence, see Fig.~\ref{fig:synth:1}.

A linear behaviour of the scaling exponents with $p$ characterizes a monofractal process. On the contrary, a non linear behaviour reveals multifractality, see Fig.~\ref{fig:synth:1}. 

From these scaling exponents, one can define the log-cumulants from the following Taylor expansion~\cite{Delour2001}:
%
%
\begin{equation}
\zeta(p)= c_{1}p-c_{2}\frac{p^2}{2!}+c_{3}\frac{p^3}{3!}\dots
\end{equation}
So, the existence of non-zero log-cumulants $c_p$ of order $p \ge 2$ indicates the multifractal nature of a process.

By taking the Legendre transform of the scaling exponents we estimate the singularity spectrum of the process~\cite{Frisch1985}:
\begin{equation}
{\cal D}(h)=\text{min}_{p}[ph-\zeta(p)] 
\end{equation}
where $h$ is called the H{\"o}lder exponent and describes the local regularity of the signal.
The singularity spectrum ${\cal D}(h)$ is related to the probability of finding the H{\"o}lder exponent $h$.

\paragraph{For a monofractal process} there is only one possible value for the H{\"o}lder exponent $h$, which is noted $\mathcal{H}$, the Hurst exponent.
%
The scaling exponent are linear in $p$ (see Fig.~\ref{fig:synth:1}): $\zeta(p)=\mathcal{H}p$ so only the first log-cumulant $c_1$ is non zero: $c_1=\mathcal{H}$.
In that case, the scale invariance implies the following relation between the probability distributions $p_{\delta_\tau X}$ and $p_{\delta_{\tau_0} X}$ of the increments of scales $\tau$ and $\tau_0$:
$$
p_{\delta_\tau X}(\delta_\tau X) = \left(\frac{\tau_0}{\tau}\right)^{\cal H} p_{\delta_{\tau_0} X}
\left(\left(\frac{\tau_0}{\tau}\right)^{\cal H}\delta_\tau X\right) \,,
$$
so:
\begin{equation}
H(\delta_\tau X) = H(\delta_{\tau_0} X) + {\cal H} \log (\tau/\tau_0) \,.
\label{eq:entropy:monofractal}
\end{equation}
This relation is valid for all couple of scales ($\tau, \tau_0$), and although there is no integral scale $T$ in a mono-fractal description, we further note the reference scale $\tau_0=T$.
Following K41, we set ${\cal H}=c_1=1/3$ to model turbulence, although this is not very satisfying for larger $p$, see Fig.~\ref{fig:synth:1}.

\paragraph{Intermittent log-normal model} for turbulence was introduced by Kolmogorov and Oboukhov in 1962. It was the first intermittent model of turbulence, with the following scaling exponents:
$ \zeta(p)= c_1 p - c_2 \frac{p^2}{2}$.

The non-linear dependence of the scaling exponents in $p$ indicates the multifractal nature of the model, which is quantified by $c_2$. All log-cumulants $c_p$ of order $p\ge 2$ are zero.
Its singularity spectrum is:

\begin{equation}
{\cal D}(h)=1-\frac{(h-c_1)^2}{2c_2} \,.
\end{equation}

This multifractal process offers a satisfying representation of the scaling exponents of Turbulence for $c_2=0,025$ and $c_1=1/3 + 3/2 c_2 = 0.37$ (see Fig.~\ref{fig:synth:1}).

\paragraph{Intermittent log-Poisson model} was introduced by She and leveque~\cite{She1994}. 
This heuristic model leads to scaling exponents of the form $\zeta(p)=-\gamma p - \lambda(\beta^{p}-1)$.
It has later been interpreted as a log-Poisson model with a singularity spectrum
\begin{equation}
{\cal D}(h)=1-\lambda+ \frac{h-\gamma}{\log(\beta)} \left( \log \left( \frac{h-\gamma}{-\lambda \log(\beta)} \right)-1 \right) \,.
\end{equation}

The corresponding log-cumulants are:

\begin{align}
 c_1&=\gamma + \lambda \log(\beta)  \label{eq:LP:def1}\\
 c_m&=\lambda \log(\beta)^{m}, \quad m\geq2 \label{eq:LP:def2}
\end{align}

This model imposes $\lambda=2$, $\beta=\left(\frac{2}{3}\right)^{(1/3)}$ and $\gamma=-1/9$~\cite{She1994}, and it describes the scaling exponents $\zeta(p)$ as satisfyingly as the log-normal model does (see Fig.~\ref{fig:synth:1}).

\subsection{Synthetic processes}

We now briefly present the different processes that we have numerically generated, according to the above prescriptions.


\paragraph{Fractional Brownian motion}
is the only scale-invariant process with Gaussian statistics and stationary increments. This monofractal process was introduced by Kolmogorov~\cite{Kolmogorov1941a} and studied by 
Mandelbrot \cite{M68}. 
%
 %
The Hurst exponent $\mathcal{H}=1/3$ and $\sigma_0$ (the variance at $t=0$) define completely the process.
Fractional Brownian motion exhibits a $5/3$ scaling, identical to the one of the energy in the inertial region of turbulence, in agreement with K41~\cite{Kolmogorov1941a}. We use the procedure presented by Helgason to synthesize fBm~\cite{Helgason2011a}.

\paragraph{log-Normal multifractal processes}
We use two different synthetic processes with log-normal statistics: a Random Wavelet Cascade (RWC)~\cite{Leonarduzzi2014,Arneodo1998} and a Multifractal Random Walk (MRW)~\cite{Bacry2001}.
Multifractality requires the existence of an integral scale $T$, from or towards which the pdf evolves.
For both processes, the synthesis we use imposes the integral scale $T$ to be equal to the size of the signal.

\paragraph{log-Poisson multifractal process}
We use a RWC with log-Poisson statistics~\cite{Arneodo1998}.
Again, our synthesis fixes the integral scale $T$ to the size of the generated signal.

\paragraph{Classical multifractal analysis}
offers a way to estimate the log-cumulants $c_1$ and $c_2$, but fails to estimate $c_3$ and higher order log-cumulants. It can therefore be interpreted as projecting the different models onto their 
log-normal approximation, with varying $(c_1, c_2)$. For example, the multifractal analysis of a realistic log-Poisson model of turbulence leads the couple of values given in table~\ref{table:1}, and no 
additional higher order log-cumulant.
As a consequence, such an analysis is not able to discriminate which process --- log-normal or log-Poisson ---better represents Turbulence.
For this reason, we compute in the next section the KL divergence $D_\tau$ which takes into account all moments of the pdf of increments, and hence higher order log-cumulants~\cite{Venugopal2006}, in order to obtain a finer analysis of the inertial domain of turbulence.

\subsection{Results}

\begin{figure}
\centering
\includegraphics[width=\linewidth]{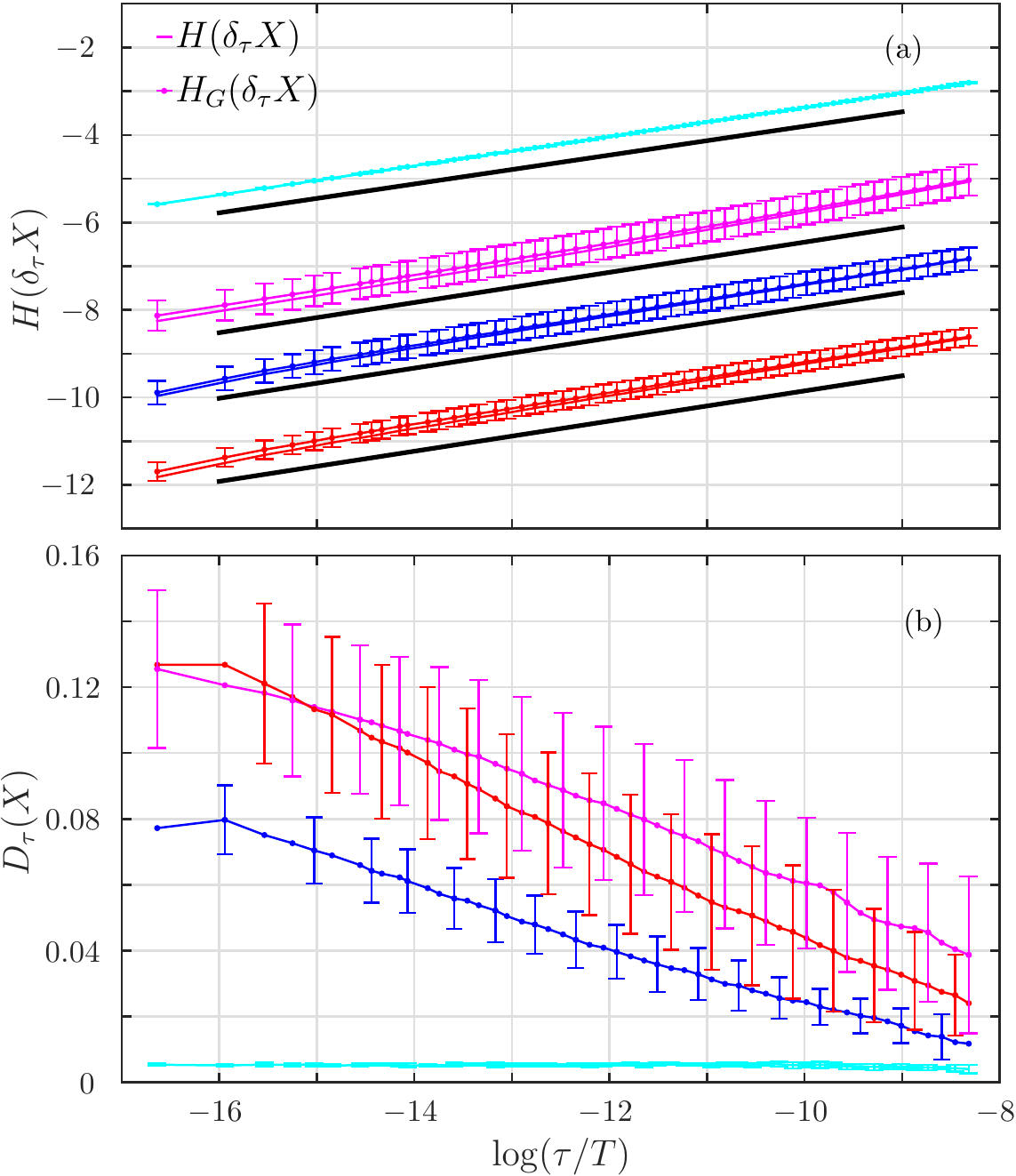}
\caption{a) Entropy $H$ and entropy under Gaussian hypothesis $H_{\rm G}$ for mono- and multifractal processes. Black lines are the theoretical slope of information in the inertial range of turbulence. 
b) Distance $D_\tau(X) = H_{\rm G}(\delta_\tau X)-H(\delta_\tau X)$ from Gaussianity.
Four different models are used: fBm (cyan), MRW (magenta), log-normal RWC (blue) and log-Poisson RWC (red).}
\label{fig:turb:1}
\end{figure}

In figure \ref{fig:turb:1}(a) we plot for the four synthetic signals the entropy $H(\delta_{\tau}X)$ as a function of $\log(\tau/T)$, the log of the scale. We also plot the entropy under Gaussian hypothesis,  $H_{\rm G}(\delta_{\tau}X)$, but it is undistinguishable from $H(\delta_{\tau}X)$.

For any process, the entropy under Gaussian hypothesis $H_{\rm G}$ is computed using eq.(\ref{eq:HG2}). It involves the second order moment $S_2(\tau)$ only, which we express using eq.~(\ref{eq:structure_function}) as
$$
S_{2}(\tau) = \sigma_\tau^2 = \sigma_{T}^2 \left( \frac{\tau}{T} \right) ^{\zeta(2)}\,.
$$
We then obtain the dependence of $H_{\rm G}$ on the scale $\tau$:
\begin{equation}
H_{\rm G}(\delta_\tau X)= H_{\rm G}(\delta_{T}X) + \frac{\zeta(2)}{2}\log\left(\tau/T\right) \,.
\label{eq:Gaussian:entropy}
\end{equation}

In figure \ref{fig:turb:1}(a), we observe that the slope of the curves, which should be $\frac{\zeta(2)}{2}$ is very similar for all processes:
we report in table~\ref{table:1} the different values we measured, and compare them to the prescribed value (1/3 for fBM and 0.345 for all three multifractal processes). The distribution of information along the scales for the four different models is in agreement with the prescribed Kolmogorov $5/3$ law~\cite{Granero16}.

\begin{table}[h] 
\begin{center}
\resizebox{\linewidth}{!}{
\begin{tabular}{|l|c|c|c|r|} \hline
                                            & fBm              & MRW            & log-N          & log-P \\ \hline
$c_1$                                       & $1/3$            & 0.370           & 0.370           & 0.381 \\ \hline
$c_2$                                       & 0                & 0.025          & 0.025          & 0.036 \\ \hline
$\zeta(2)/2 $                                   & $1/3$            & 0.345          & 0.345          & 0.345 \\ \hline
$\hat{c}_1$ & 0.333& 0.42 & 0.372 & 0.382 \\ \hline
$\hat{c}_2$ & $1e^{-4}$ & 0.038 & 0.026 &0.035\\ \hline
$\hat{\zeta}(2)/2$ & 0.332 & 0.363 & 0.353 & 0.356\\ \hline
$\Delta_{\log(\tau)} H_G(\delta_{\tau}X)$ & 0.33$\pm0.01$    & 0.37$\pm0.01$  & 0.35$\pm0.01$  & 0.35$\pm0.01$ \\ \hline
\end{tabular}
}
\caption{The first three lines indicate the values of parameters ($c_1$ and $c_2$ and hence $\zeta(2)$) used in the generation. Estimates $\hat{c}_1$, $\hat{c}_2$ and $\hat{\zeta}(2)$ are obtained by classical multifractal analysis.
Last line reports the slopes $\Delta_{\log(\tau)} H_G(\delta_{\tau}X)$ of the entropy $H_G(\delta_{\tau}X)$ as a function of $\log(\tau/T)$, for the four different models, which according to eq.(\ref{eq:Gaussian:entropy}) provides another estimate of $\zeta(2)/2$.}
\label{table:1}
\end{center}
\end{table}

Up to this point, looking at the entropies, the four models cannot be distinguished in the inertial domain.
In figure \ref{fig:turb:1}(b) we plot the Kullback-Leibler divergence $D_{\tau}(X)$ as a function of $\log(\tau/T) = \log(l/L)$, for scales ranging from $\tau/T=1/2^{24}$ to $\tau/T=4096/2^{24}$ where the integral scale is $T=2^{24}$.

For a monofractal process, the entropy is given by eq.(\ref{eq:entropy:monofractal}), and the entropy under Gaussian hypothesis is given by eq.(\ref{eq:Gaussian:entropy}) with ${\cal H}=\zeta(2)/2$, so $D_\tau(X) = H_{\rm G}(\delta_{\tau} X) - H(\delta_{\tau} X)$ is constant and does not depend on the scale $\tau$.
If the monofractal process has Gaussian statistics ---~which defines the fBm~--- $D_\tau(X) =0$ by construction.
Looking at Fig.~ \ref{fig:turb:1}(b), $D_\tau$ for the fBm is not exactly zero; this is due to the bias in the estimation of $H(\delta_{\tau}X)$ and $H_{G}(\delta_{\tau}X)$. This bias is constant across scales, because our procedure was built to use a constant number of points in the range of $\tau$ we use.

For the three multifractal processes, $D_{\tau}(X)$ decreases monotonically when $\tau$ increases, and tends to zero when the scale tends to the integral scale.
So in the three multifractal models, the pdfs of the increments deform into a Gaussian pdf when approaching the integral scale.
Moreover, in Fig.~\ref{fig:turb:1}(b), we observe that the three processes, which indeed have different statistics, do not converge to zero in the same way. 
The distance from Gaussianity $D_\tau$, by involving all the moments of the probability distributions, is able to reveal fine differences between processes.

The synthetic processes used above are good representations of the inertial range only.
They do not properly take into account either the dissipative nor the integral scales.
Nevertheless, the synthesis imposes an effective integral scale that corresponds to the size of the generated signal.
In order to study more precisely the deformation of the pdfs at large scale, we now turn to descriptions that explicitly involve the integral scale.

\subsection{Phenomenological model : the propagator formalism} 

First introduced by Castaing~\cite{Castaing1990}, the propagator formalism describes the statistics of the Eulerian velocity increment $\delta_{l}v$ as identical, in the probabilistic sense, to the statistics of the product of two random variables: the large scale fluctuations $\sigma_L \delta$ and
the propagator $(l/L)^h$. The large scale fluctuations are supposed Gaussian, with standard deviation $\sigma_L$, and $\delta$ is therefore a Gaussian variable with unit variance.
The propagator deforms the large-scale statistics when the scale $l$ is reduced below the integral scale $L$.
In the simple situation where no dissipative scale is taken into account, and where the propagator is supposed independent of large scale statistics, one can write formally the pdf of the Eulerian velocity increments $\delta_{l}v=\sigma_L(l/L)^{h}\delta$ as~\cite{Chevillard2012}:
\begin{equation}\label{eq:propagator}
p_{\delta_{l}v}(\delta_{l}v)
=\int_{-\infty}^{\infty} \frac{1}{\sigma_L}\left(\frac{l}{L}\right)^{-h}
\mathcal{P}_{\delta}\left[\frac{\delta_{l}v}{\sigma_L}\left(\frac{l}{L}\right)^{-h}\right]
\mathcal{P}_{h}[h] 
dh
\end{equation}
where $h$ is the H{\"o}lder exponent.
We have noted $\mathcal{P}_{\delta}(\delta)$ and $\mathcal{P}_{h}(h)$ the probabilities of the independent random variables $\delta$ and $h$. 
The pdf $\mathcal{P}_{h}(h)$ depends only on the singularity spectrum ${\cal D}(h)$.
See \cite{Chevillard2012} for a detailed explanation.

\begin{figure}
\centering
\includegraphics[width=\linewidth]{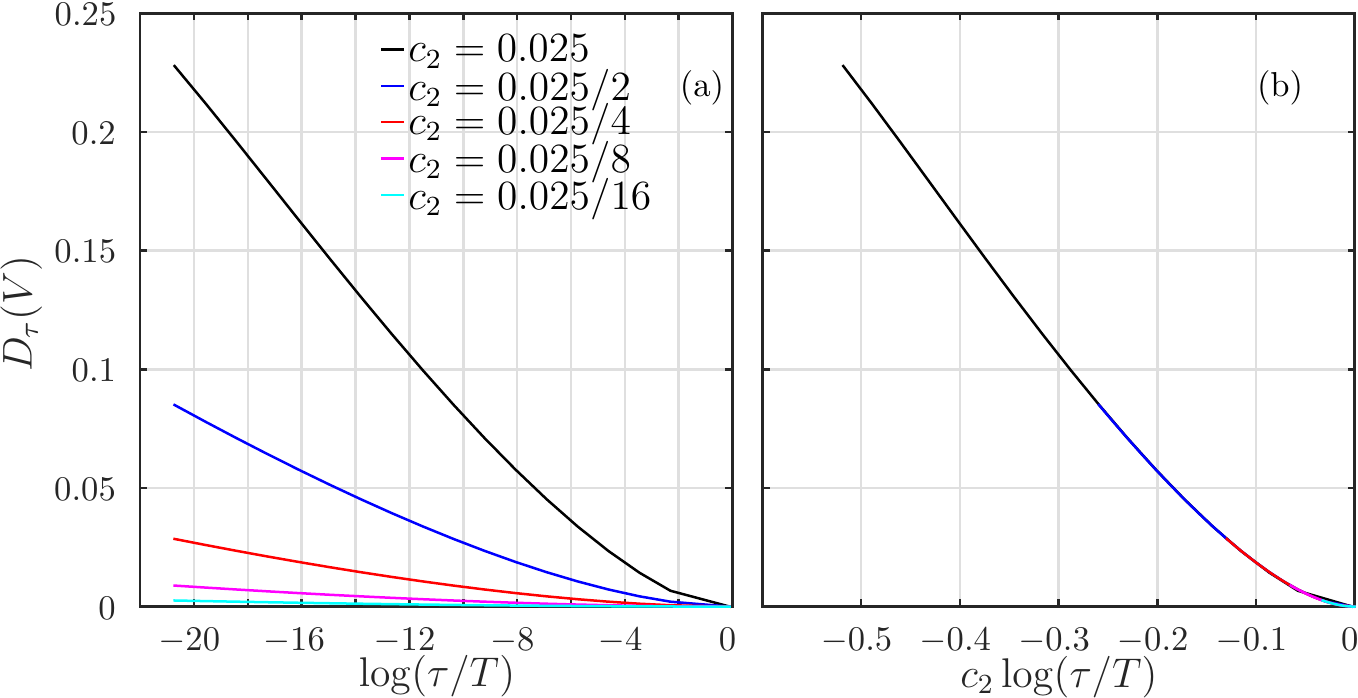}
\caption{Kullback-Leibler divergence $D_\tau$ for the log-normal propagator model, for varying values of the log-cumulant $c_2$, as a function of $\log(\tau/T)=\log(l/L)$ (a) or as a function of $c_2 \log(\tau/T)$ (b).}
\label{fig:propagator:ln}
\end{figure}

We integrate numerically eq.(\ref{eq:propagator}) to get the pdf of the increments $\delta_{l}v$, and then compute the KL divergence $D_\tau$ for several singularity spectra, either log-normal or log-Poisson.

\paragraph{log-normal model}. 
We varied the value of the log-cumulant $c_1$ and didn't observed any dependence of $D_\tau$ on $c_1$. 
On the contrary, varying $c_2$ strongly changes the convergence.
Results are presented in Fig.~\ref{fig:propagator:ln}(a).
We observe and report in Fig.~\ref{fig:propagator:ln}(b) that curves for different values of $c_2$ can be collapsed into a single curve when plotted as a function of $c_2 \log(\tau/T)=c_2\log(l/L)$.

To understand this scaling behavior, we performed a saddle-node expansion of expression (\ref{eq:propagator}) in the log-nomal case, and obtained the following simplified expression for the pdf of the normalized increments $y=\delta_{l}v/\sigma_l$ at scale $l$:
\begin{equation}
p_y(y) = \frac{e^{\frac{3}{2}c_2 x}}{\sqrt{2\pi}}\frac{\displaystyle  e^{-\frac{2W+W^2}{8c_2 x}}}{\sqrt{1+W}}
\label{eq:p_sn_theo}
\end{equation}
where we have noted $x\equiv-\log(l/L)$ the logarithmic scale, and $W$ the value of the Lambert W-function of argument $2c_2 x y^2 e^{4c_2 x}$. Eq.(\ref{eq:p_sn_theo}) is a non-Gaussian pdf which converges to the Gaussian pdf of variance $\sigma_L^2$ when $x\rightarrow 0$. 
From eq.(\ref{eq:p_sn_theo}), the pdf of the increments only depends on $\log(l/L)=\log(\tau/T)$ and $c_2$ via the product $c_2 \log(l/L)$. As a consequence, the entropy of the increments depends on the scale $l$ as $c_2 \ln{l/L}$ only. This implies that the KL divergence $D_\tau$ for the log-normal  process has the scaling observed in Fig.~\ref{fig:propagator:ln}(b).

\begin{figure}
\centering
\includegraphics[width=\linewidth]{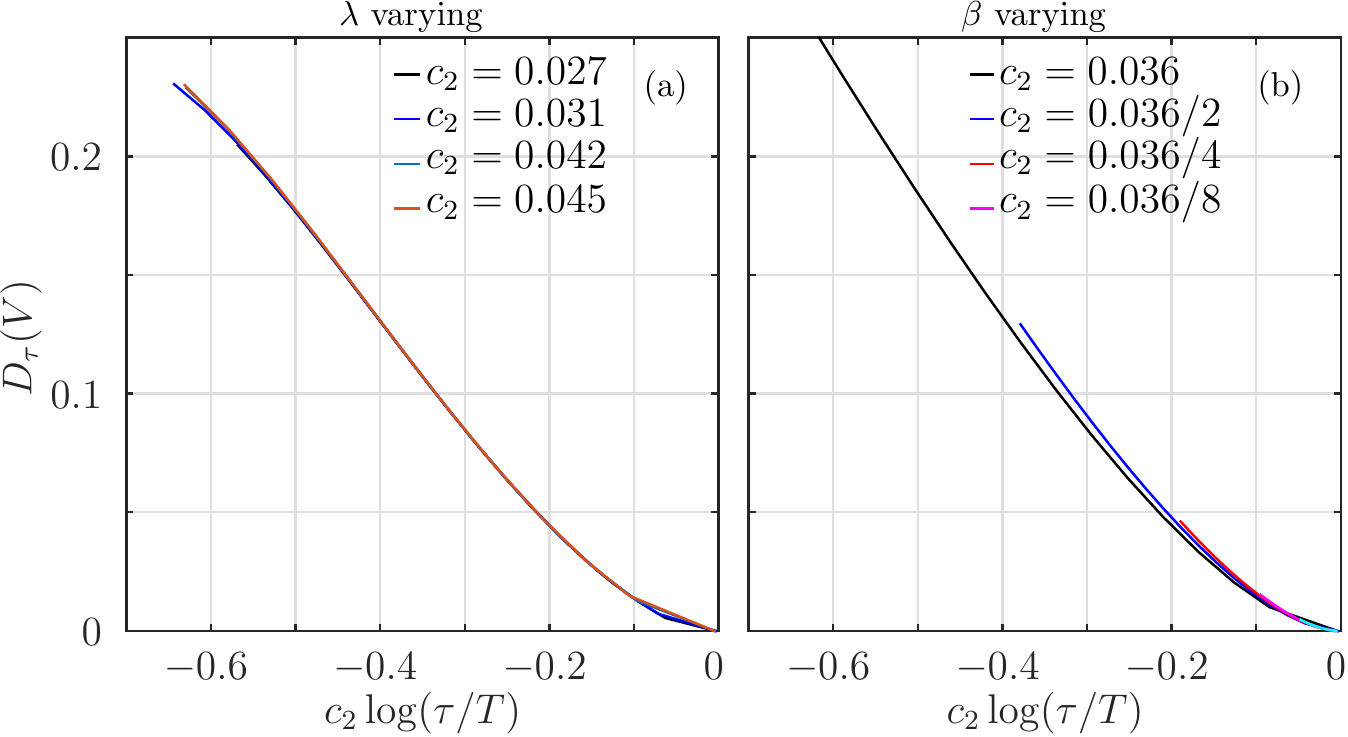}
\caption{Kullback-Leibler divergence $D_\tau$ as a function of $c_2 \log(\tau/T)$ for the log-Poisson propagator model, for varying values of $\lambda$ (a) and $\beta$ (b).}
\label{fig:propagator:lP}
\end{figure}

\paragraph{log-Poisson model}
We varied independently $\gamma$, $\lambda$ and $\beta$.
We didn't observed any change of $D_\tau$ when $\gamma$ was varied. This can be understood as $\gamma$ only changes the value of $c_1$ (see eq.(\ref{eq:LP:def1})), which does not impact $D_\tau$.
Varying $\lambda$ changes the convergence, as this amounts to change $c_2$ (see eq.(\ref{eq:LP:def2})), but we observe again that $D_\tau$ depends only on $c_2 \log(\tau/T)$, see Fig.~\ref{fig:propagator:lP}(a). This can be understood by noting that all log-cumulants are linear in $\lambda$; thus  varying $\lambda$ amounts to a change of $c_2$ while keeping higher order cumulants within the same ratio.
On the contrary, varying $\beta$ has more impact on the convergence, and the re-scaling in $c_2 \log(\tau/T)$ is then not perfect, albeit still relevant, see Fig.~\ref{fig:propagator:lP}(b). This can be understood by noting that changing $\beta$ will not only change $c_2$, but also the ratio of all higher order cumulants.

\paragraph{Comparison between models and Turbulence data}

\begin{figure}
\centering
\includegraphics[width=\linewidth]{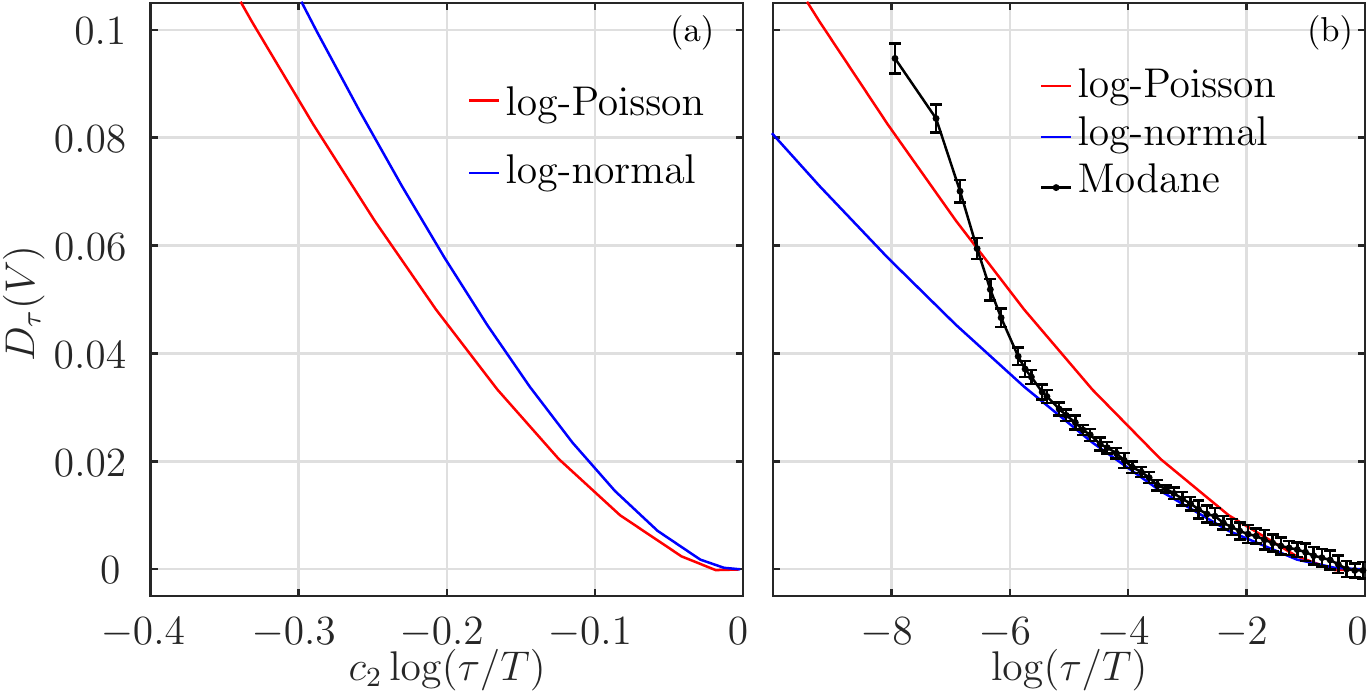}
\caption{(a) Comparison of the evolution of $D_\tau$ as a function of the re-scaled scale $c_2\log(\tau/T)$ for the log-normal ($c_2=0.025$) and log-Poisson ($c_2=0.036$) models of turbulence. (b) Comparison of the two models with Modane experimental data, as a function of $\log(\tau/T)$.}
\label{fig:comparison}
\end{figure}

Amongst open questions regarding statistical descriptions of Eulerian Turbulence is the choice of a log-normal or log-Poisson modelling of its multifractal nature. We of course want to adress this issue, and 
propose in Fig.~\ref{fig:comparison}(a) a comparison of the KL divergence of the two models in function of $c_2\log(\tau/T)$. We have used in each case the $c_2$ value expected for turbulence 
(see table~\ref{table:1}). 
The rescaling in $c_2 \log(\tau/T)$ ---~which absorbs most, if not all, the dependence of $D_\tau$ on $c_2$~---- allows a direct comparison of models. The obtained curves are clearly different, which probably results from the presence of higher order log-cumulants $c_p$, $p>2$ in the log-Poisson propagator.

In Fig.\ref{fig:comparison}b, we compare the two models with Modane experimental data. 
To do so, we remove the bias estimated from fBm measurements, see Fig.~\ref{fig:turb:1}(b).
As the $c_2$ value for turbulence is {\em a priori} unknown, we do not rescale the x-axis with $c_2$. Let us remark though that the $c_2$ value we have used in the log-normal propagator ($c_2=0.025$) is exactly the one measured on the experimental data, using multifractal analysis~\cite{Chevillard2012}.
Our results show a (much) better agreement of the experimental data with the log-normal model.
Although this may be due to the very appropriate choice of $c_2$, the log-Poisson model does not allow such a choice and fixes all the log-cumulants~\cite{She1994}.
As a consequence, we can state that the deformation of the experimental velocity increments pdf is better modelled by a multiplicative cascade with log-normal multipliers.

In the dissipative range, {\em i.e. ,} for smaller scales, we observe a rapid increase of $D_\tau$ in the experiments, very different from the predictions of the two models. This is expected as both models used here do not incorporate any modelling of the dissipative scales.

\section{Discussion and Conclusions}

We have measured the Shannon entropy of the Eulerian turbulent velocity increments, and studied its dependence on the scale of the increments. We have recovered three different behaviors in the integral, inertial and dissipative domains, in perfect agreement with the classical analysis using the power spectrum. In particular, in the inertial range, a scaling law for the entropy is observed, reminiscent of K41 theory, similar to what was earlier reported for another Information Theory quantity~\cite{Granero16}.
A closer look at the entropy, and especially a comparison with its Gaussian approximation, which only takes into account the variance of the signal --~exactly as the PSD does~--- allows a much finer description and in particular a measure of intermittency, as introduced in KO62.

We have proposed a quantitative measure of intermittency.
Although some quantities were already used as an intermittency coefficient, most, if not all, were ratios of structure functions~\cite{Frisch1995}, and as such, they were depending on the chosen ratio: flatness, hyper-flatness~\cite{F.Anselmet1984} or higher order ratios. 
We interpret intermittency as the distance from Gaussianity, and measure it as $D_\tau$, the Kullback-Leibler divergence between the complete pdf $p(\delta_{\tau}V)$ and its Gaussian approximation $p_{\rm G}(\delta_{\tau}V)$; the first involves all the statistical moments while the second one only depends on the  variance. 
Our measure of intermittency, by comparing complete pdfs, takes into account all the moments of the distributions, which leaves no room for ambiguity on the choice of the moments. 

 We have checked the robustness of our approach by analyzing several experimental datasets, from two different experimental setups, and with varying Reynolds numbers.

The quantity $D_{\tau}$ is not only able to measure intermittency in turbulence, but also to discriminate very easily monofractal from multifractal processes. 
Furthermore, the evolution of $D_\tau$ with the scale depends on the process: this provides a much more precise characterization of the process than the bare set of log-cumulant values $(c_1,c_2)$ given by a regular multifractal analysis.
This may be exploited to discriminate log-Poisson from log-normal models of intermittency in turbulence. 


We have investigated the dependence of $D_\tau$ on the log-cumulants.
$D_\tau$ does not depend on $c_1$ and we have captured its dependence on $c_2$, and especially how it affects the convergence to 0 at large scales.
%
Because $D_\tau$ appears to depend on $c_2 \log(\tau/T)$, we can state that the speed of the deformation of the pdf, starting from a Gaussian at large scale $L$, depends on $c_2$.
For a given scale $l/L$, or equivalently $\tau/T$, the deformation of the pdf, and hence the intermittency, is an increasing function of $c_2$.
Conversely, for a fixed value of $c_2$, the influence --- or reminiscence--- of the integral scale persists down to scales $l/L$ smaller and smaller when $c_2$ is reduced.
Because the typical $c_2$ of turbulence is small, the influence of the integral scale persists in the inertial domain, down to the dissipative domain, unless the Reynolds number tends to arbitrarily large values.
We have shown that $D_\tau$ depends on higher order log-cumulants $c_p$, for $p>2$, by looking at the special case of log-Poisson statistics (Fig.~\ref{fig:propagator:lP}). The dependence seems weak, but is nevertheless present, and could be exploited.

Although we have put a strong emphasis on turbulence, we want to point out that our approach is extremely general and should find successful applications in many other fields.
It should prove particularly interesting for non-Gaussian processes, the most common in Nature and Society. Any multifractal process, or process that may be considered multifractal in some range of scales can be analyzed with $D_\tau$. The local intermittency measure that $D_\tau$ provides can be used to characterize the process at any scale.

\begin{acknowledgments}
The authors wish to thank L. Chevillard for stimulating discussions.
This work was supported by the LABEX iMUST (ANR-10-LABX-0064) of Universit\'e de Lyon, within the program "Investissements d'Avenir" (ANR-11-IDEX-0007) operated by the French National Research Agency (ANR).
\end{acknowledgments}

\bibliographystyle{unsrt}
\bibliography{biblio}

\end{document}